\documentclass[aps,superscriptaddress,twocolumn,driverfallback=dvips]{revtex4}
\usepackage{feynmp}
\usepackage{amsmath}
\usepackage{amssymb}
\usepackage{epsfig}
\usepackage{graphicx}
\usepackage{subfigure}
\usepackage{hyperref}
\usepackage{bbm}
\usepackage{epsfig}
\usepackage{color}
\usepackage{float}
\usepackage{mhchem}
\usepackage{epstopdf}

\newcommand{\Rmnum}[1]{\expandafter\@slowromancap\romannumeral #1@}
\allowdisplaybreaks[3]

\begin{document}

\title{Direct Urca process facilitated by non-Fermi-liquid quantum criticality}

\author{Hao-Fu Zhu}
\affiliation{Department of Astronomy, University of Science and
Technology of China, Hefei, Anhui 230026, China}\affiliation{School
of Astronomy and Space Science, University of Science and Technology
of China, Hefei, Anhui 230026, China}
\author{Guo-Zhu Liu}
\altaffiliation{Corresponding author: gzliu@ustc.edu.cn}
\affiliation{Department of Modern Physics, University of Science and
Technology of China, Hefei, Anhui 230026, China}

\begin{abstract}
Conventional Fermi-liquid theories of dense nuclear matter predict that the direct Urca process requires a proton fraction above a threshold of $Y_p \approx 0.11$--$0.15$, which is rarely reached in low-mass neutron stars. Here we show that this stringent kinematic constraint is drastically relaxed by the non-Fermi liquid behavior arising from superfluid and superconducting quantum criticality in neutron stars. In the non-Fermi-liquid regimes, nucleons have a smooth momentum occupation without a sharp Fermi surface, and their chemical potentials develop a logarithmic temperature dependence. These intrinsic features fundamentally alter the standard $\beta$-equilibrium relation. Consequently, direct Urca neutrino emission can be activated even at low nuclear densities. This modifies the long-standing Urca-threshold paradigm and indicates that direct Urca process has the opportunity to occur in low-mass neutron stars.
\end{abstract}

\maketitle


Shortly after their birth in supernova explosions, the neutron stars (NSs) possess interior temperatures as high as $\approx (10^{11}$-$10^{12})\mathrm{K}$. Within a few minutes, they lose a large amount of energy via neutrino emission. At this stage, the dominant cooling channel is the nucleonic direct Urca (nDU) process, which refers to the beta decay and lepton capture reactions among neutrons and protons ($n\rightarrow p+\ell^-+\bar{\nu}_\ell$, $p+\ell^-\rightarrow n+\nu_\ell$; $\ell\in \{e,\mu$\}) \cite{Gamow41}. Such process must conserve both the energy and momentum. As NSs further cool down to lower temperatures, these constraints can hardly be fulfilled simultaneously~\cite{Yakovlev01, Page06a, Potekhin15, Tsuruta23}. Therefore, the nDU process is strongly suppressed in ordinary NSs, operating efficiently only during the first few minutes after formation.

Prior to the 1990s, the nDU process was considered forbidden in ordinary NSs unless certain exotic degrees of freedom emerged~\cite{Bahcall65, Iwamoto80, Brown88}, except for Boguta~\cite{Boguta81}. Lattimer \emph{et al.}~\cite{Lattimer91, Page92, Pethick92} re-examined this issue and established that the nDU process can operate in nuclear matter if the proton fraction exceeds a threshold of $Y_p \approx (0.11$--$0.15$). This condition is temperature independent and could be satisfied in $\beta$-equilibrated NSs only when the central density reaches a sufficiently high value, which corresponds to a relatively large threshold NS mass $M_{\mathrm{DU}}$~\cite{Klahn06, Beznogov15, Boukari24, Xu26, Montefusco26}. The onset of nDU process produces a neutrino luminosity far exceeding that of other cooling mechanisms (e.g., modified Urca process: $n+N\to p+N+\ell^-+\bar{\nu}_\ell$, $p+N+\ell^-\to n+N+\nu_\ell$; $N\in \{n,p\}$)~\cite{Yakovlev01, Page06a, Potekhin15, Tsuruta23}, and has been applied to account for the rapid cooling observed in isolated~\cite{Marino24, Ho24} and accreting~\cite{Han17, Brown18} NS systems.

In standard picture, the triggering condition for nDU process is derived assuming an ideal, non-interacting degenerate quantum gas (DQG) of unpaired nucleons, and is determined by combining electrical neutrality, $\beta$-equilibrium, and momentum conservation. The chemical potentials of neutrons, protons, and electrons are denoted as $\mu_n$, $\mu_p$, and $\mu_e$, respectively. They are required by the $\beta$-equilibrium condition to obey the identity
\begin{equation}
\mu_{n}=\mu_{p}+\mu_{e}, \label{energyconservation}
\end{equation}
with $\mu_{\nu}=0$ for freely escaping neutrinos and the muon contribution neglected. Electrical neutrality implies that the particle densities of electrons and protons are equal, i.e., $\rho_e=\rho_p$. Since $\rho_e=k_{\mathrm{Fe}}^3/3\pi^2\hbar^3$ and $\rho_p=k_{\mathrm{Fp}}^3/3\pi^2\hbar^3$, where $k_{\mathrm{F}}$ is the Fermi momentum, this condition is equivalent to $k_{\mathrm{Fe}}=k_{\mathrm{Fp}}$. Momentum conservation dictates that $k_{\mathrm{Fe}}$, $k_{\mathrm{Fp}}$, and $k_{\mathrm{Fn}}$ must satisfy the triangle inequalities, among which the most restrictive one is
\begin{equation}
k_{\mathrm{Fe}}+k_{\mathrm{Fp}}\geq k_{\mathrm{Fn}}, \label{triangle}
\end{equation}
since $k_{\mathrm{Fn}}$ is usually much larger than $k_{\mathrm{Fe}}$ and $k_{\mathrm{Fp}}$ in NSs. This inequality leads to the neutron density bound $\rho_n \leq 8\rho_p$. Therefore, nDU processes are allowed only when the proton fraction $Y_p=\rho_p/(\rho_n+\rho_p)$ exceeds the threshold of $\approx 0.11$.

The ideal DQG assumption may break down in the presence of significant quantum many-body effects. In particular, the strongly correlated nucleons may form a non-Fermi liquid (NFL) in certain regions of cold dense NS matter~\cite{Zhu25,Zhu26}. The NFL regime exhibits several unique properties: nucleon quasiparticles are no longer well defined, the sharp Fermi surface becomes smeared, and a substantial fraction of nucleons are excited out of the Fermi sphere. Furthermore, at a fixed density, the originally constant chemical potentials $\mu_{n,p}$ can develop a singular temperature dependence. In this case, the $\beta$-equilibrium relation given by Eq.~\eqref{energyconservation} is changed to
\begin{equation}
\mu_{n}^{\mathrm{NFL}}(T)=\mu_{p}^{\mathrm{NFL}}(T)+\mu_e,
\label{eq:newbetacondition}
\end{equation}
where $\mu_{i}^{\mathrm{NFL}}(T)$ ($i=n, p$) denote the NFL-corrected chemical potentials. It is worthwhile to investigate how these NFL effects alter the standard nDU threshold.

\begin{figure}[htbp]
\centering
\includegraphics[width=\linewidth]{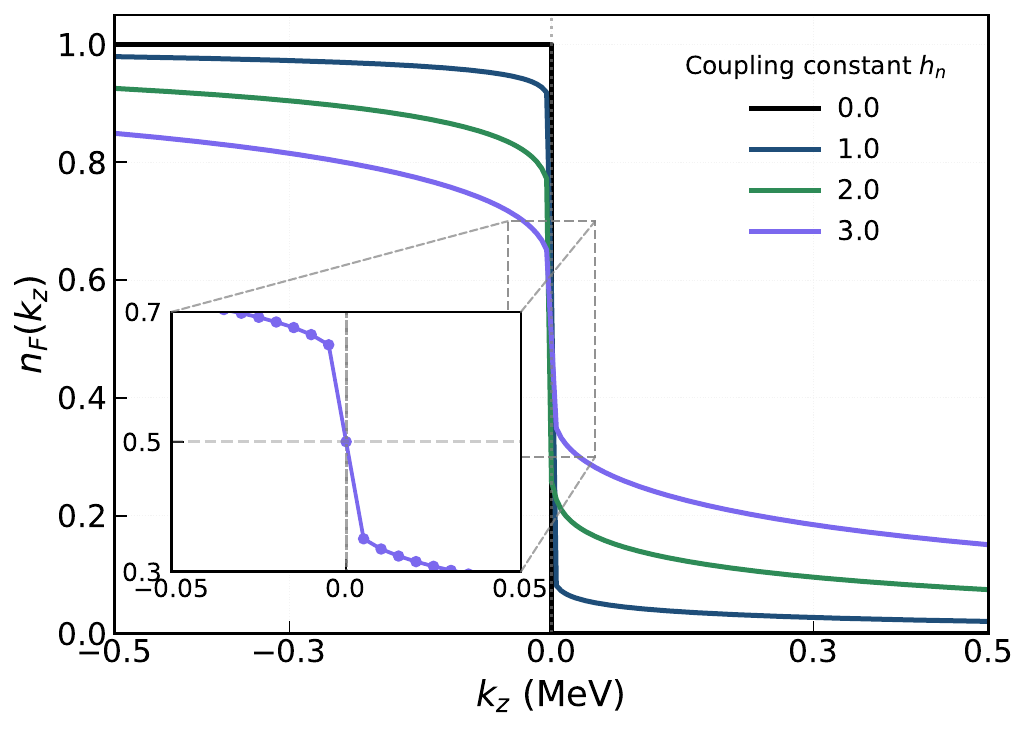}
\caption{Neutron occupation number $n_{\mathrm{F}}(k_z)$ in the $^3P_2$-wave neutron superfluid quantum critical region. The curves show $n_{\mathrm{F}}(k_z)$ at $T=0$ and $k_\perp=\sqrt{k_x^{2}+k_y^{2}}=0$ versus the longitudinal momentum $k_z$ for $h_n=0$, $1.0$, $2.0$, and $3.0$; the case $h_n=0$ corresponds to the non-interacting limit. The inset enlarges the region $|k_z|\le 0.05$~MeV and shows only the $h_n=3.0$ case. The NFL behavior manifests itself as a smeared Fermi surface.}
\label{fig:occupation}
\end{figure}

In this paper, we study the impact of NFL behavior on the triggering condition of nDU process as well as on the resulting neutrino emissivity. In our scenario, the NFL behavior originates from superfluid and superconducting quantum criticality in NSs~\cite{Zhu25,Zhu26}. After including this behavior, we find that the conventional nDU triggering condition is significantly relaxed and the nDU neutrino emission can operate even at low nuclear densities.

This work is motivated by the fact that the neutron superfluid and proton superconducting phases predicted in prior works occupy finite density regions. Extensive calculations have found three nucleon pairing phases in dense nuclear matter~\cite{Sedrakian19, Pagereview}, including $^{1}S_{0}$-wave neutron superfluid phase, $^{3}P_{2}$-wave neutron superfluid phase, and $^{1}S_{0}$-wave proton superconducting phase. Each ordered phase has a dome-shaped phase transition line with two end points. Each end point corresponds to a critical nuclear density separating ordered and disordered phases, and can be naturally identified as a quantum critical point (QCP)~\cite{Zhu25, Zhu26}. In condensed matter physics~\cite{Vojta03, Coleman05, Loehneysen07, Sachdevbook}, it is established that unusual NFL-type quantum critical phenomena emerge in the vicinity of various QCPs. Near the superfluid/superconducting QCPs, the quantum fluctuations of the associated order parameter are interacting with unpaired nucleons. As demonstrated in Ref.~\cite{Zhu25}, such an interaction leads to NFL behavior of the nucleons. The imaginary part of retarded nucleon self-energy~\cite{Zhu25}, denoted by $\Sigma_R(\omega,T)$, exhibits a linear $\omega$-dependence at $T=0$ and a linear $T$-dependence at $\omega = 0$, i.e., $\mathrm{Im}\,\Sigma_R(\omega,T) \propto \max(\omega,T)$. The quasiparticle residue $Z$, which measures the overlap between non-interacting and interacting fermion systems~\cite{Varma, Giuliani}, is defined as
\begin{eqnarray}
Z = \lim_{\omega\rightarrow 0} \Big[1-\frac{\partial}{\partial
\omega}\mathrm{Re}\Sigma_R(\omega)\Big]^{-1},\label{eq:residue}
\end{eqnarray}
where the real part $\mathrm{Re}\Sigma_R(\omega)$ is linked to the imaginary part via the Kramers-Kronig relation. It can be verified (see Supplementary materials) that the linear $\omega$-dependence results in a vanishing quasiparticle residue, i.e., $Z=0$. This is a clear signature of the breakdown of Fermi liquid (FL) description. In contrast, the residue $Z=1$ in a DQG and $0<Z<1$ in an ordinary FL, implying the existence of long-lived quasiparticles.

The vanishing of $Z$ is clearly manifested in the nucleon momentum occupation $n_{\mathrm{F}}(\mathbf{k})$. In a DQG, $n_{\mathrm{F}}^{\mathrm{DQG}}(\mathbf{k}) = \Theta(|\mathbf{k}|-k_{\mathrm{F}})$, so only nucleons on the Fermi surface participate in the nDU process. A FL is similar to a DQG in that $Z$ remains finite and $n_{\mathrm{F}}^{\mathrm{FL}}(\mathbf{k})$ still displays a sharp Fermi surface~\cite{Migdal67, Dong16}. The NFL regime is entirely different from DQG/FL because the vanishing of $Z$ implies that $n_{\mathrm{F}}^{\mathrm{NFL}}(\mathbf{k})$ varies continuously with nucleon momentum without any abrupt change. This is visualized in Fig.~\ref{fig:occupation}, which shows the zero-$T$ occupation $n_{\mathrm{F}}(k_z)$ obtained after including NFL corrections (The corresponding spectral function $A(\omega,\mathbf{k})$ is shown in Fig.~\ref{fig:spectral} of the Supplementary materials). One can see that the NFL effects smear the step in $n_{\mathrm{F}}$ into a continuous crossover that becomes progressively less steep with increasing $h_n$ (the coupling strength defined below). Therefore, in the NFL regime all nucleons have the opportunity to participate in nDU processes, and the threshold condition is expected to be drastically relaxed.

We now incorporate the NFL effects and explore their influence on the proton-fraction threshold. In a degenerate dense medium, the quasiparticle excitation spectrum of nucleons is $\xi_{i}(\mathbf{k}) = \sqrt{c^{2}\mathbf{k}^2 + c^{4}{M^{\ast}_{i}}^{2}} - \mu_{i}$, where $c$ denotes the speed of light. For each nucleon species $i$, $M^{\ast}_i$ is the effective mass, which reduces to the bare mass $M_i$ in the noninteracting limit, while the bare chemical potential is given by $\mu_{i} = \sqrt{c^{2}k^{2}_{\mathrm{F}i} + c^{4}M^{\ast 2}_{i}}$. The background potential, which is nearly momentum-independent near the Fermi surface, only shifts $\mu_i$ and is not written explicitly. It is convenient to define the nucleon velocity~\cite{Zhu25} as
\begin{eqnarray}
v_{\mathrm{F}i} = c^{2}k_{\mathrm{F}i}/\mu_{i}.
\label{eq:generalvelocity}
\end{eqnarray}
Near $k_{\mathrm{F}i}$, the spectrum can be approximated as
\begin{eqnarray}
\xi_{i}(\mathbf{k}) \approx
\frac{v_{\mathrm{F}i}}{2 k_{\mathrm{F}i}}k_{\perp}^{2}
+v_{\mathrm{F}i}k_{z}.
\label{eq:lineardisperson}
\end{eqnarray}
The Fermi velocities $v^{~}_{\mathrm{Fn}}$ and $v^{~}_{\mathrm{Fp}}$ are renormalized by the quantum fluctuations of superfluid and superconducting order parameters, respectively. After performing a renormalization group (RG) analysis~\cite{Zhu25}, we have obtained the following renormalized velocities
\begin{eqnarray}
v^{~}_{\mathrm{Fi}}(l) = \frac{a_i\pi^{2}
v^{2}_{\mathrm{Fi}}}{h^{2}_{i}l + a_i\pi^2v^{~}_{\mathrm{Fi}}}.
\label{protonvelocity}
\end{eqnarray}
Here, $a_n=8/5$ and $a_p=4$. The RG flow parameter $l$ is defined by the scaling relation $\omega(l)=\omega_0 e^{-2l}$, where $\omega_0$ is the ultraviolet cutoff, and $\omega(l)$ represents the energy scale at RG scale $l$. At $l=0$, we recover the highest-energy ultraviolet scale $\omega(l=0)=\omega_{0}$. Increasing $l$ corresponds to the RG flow toward lower energies. Taking $l\to\infty$ yields the infrared long-wavelength limit $\omega(l\to\infty)=0$. Following Ref.~\cite{Zhu26}, we introduce two effective dimensionless coupling constants $h_n$ and $h_p$ to describe the interaction between unpaired nucleons and the associated quantum fluctuations. Their microscopic values depend on several factors~\cite{Zhu26}, such as the local density, temperature, and magnetic field, and cannot be reliably computed. We therefore absorb this dependence into $h_n$ and $h_p$, which are subject to observational constraints.

From Eq.(\ref{eq:generalvelocity}) and Eq.(\ref{protonvelocity}), we learn that the ratio $k_{\mathrm{F}i}/\mu_i$ becomes $l$-dependent. Our strategy here is to fix $k_{\mathrm{F}i}$ while requiring $\mu_{i}$ to flow with varying $l$. At fixed particle densities and temperature, the generalized Luttinger theorem~\cite{Luttinger60,Dzyaloshinskii03,Georges12} ensures that the NFL corrections do not shift the Fermi volume, so the relation $\rho_i \propto k_{\mathrm{F}i}^3$ holds and hence $k_{\mathrm{F}i}$ are fixed (see Supplementary materials for details). The NFL corrections to the Fermi velocities are then absorbed entirely into the chemical potentials, which develop the following $l$-dependence:
\begin{eqnarray}
\mu_{i}(l) &=& \frac{c^2k^{~}_{\mathrm{Fi}}(h^{2}_{i}l +
a_i\pi^{2} v_{\mathrm{Fi}})}{a_{i}\pi^{2} v^{2}_{\mathrm{Fi}}}. \label{NFLpchl}
\end{eqnarray}
In the limit $l\to 0$, these expressions reduce to the bare chemical potentials $\mu_i(l=0)=\mu_i$.

Within the RG framework, the dependence on $l$ (or the energy scale $\omega$) can be mapped onto the $T$-dependence via the correspondence $\hbar\omega \sim k_B T$, which identifies the RG energy scale with the thermal energy. Since $\omega(l)=\omega_0 e^{-2l}$, an arbitrary temperature $T$ corresponds to the relation $T(l)=T_0 e^{-2l}$, where $T_0 = \omega_0/k_B$ is the ultraviolet cutoff in temperature units. For typical NS conditions, $T_0$ is of the order of $10^{11}$~K~\cite{Burrows86, Fisher10, Hudepohl10}, corresponding to the temperature scale of the post-formation epoch. It then follows that
\begin{eqnarray}
\mu^{\mathrm{NFL}}_{i}(T) &=& \mu_{i} + \frac{h^{2}_{n}
\mu^{2}_{i}}{2a_i\pi^{2} c^2k^{~}_{\mathrm{Fi}}}
\ln\Big(\frac{T_{0}}{T}\Big).\label{NFLpchT}
\end{eqnarray}
These logarithmic corrections are a hallmark of NFL physics, in contrast to the $\sim T^2$ corrections in both a DQG and a FL. For $T$ slightly below $T_0$, the logarithmic term is negligible since $\ln(T_0/T)\approx 0$. As $T$ decreases further, the logarithmic corrections are enhanced and eventually dominates over the bare chemical potentials at sufficiently low temperatures. Similar NFL logarithmic corrections have also been identified in ultradegenerate electron and quark matter~\cite{Gerhold04}.

The electrons are unaffected by nucleon NFL corrections and thus are treated as an ideal quantum gas. Their chemical potential $\mu_{e}$ depends on $T$ via the Sommerfeld expansion $\mu_{e}(T)=\mu_{e}[1-\frac{\pi^{2}}{12}(\frac{k_{B}T}{\mu_{e}})^{2}]$. The $\sim T^{2}$ correction is negligible in the limit $k_{B}T\ll\mu_{e}$. Therefore, the $\beta$-equilibrium condition is modified solely through the NFL-type logarithmic corrections to nucleon chemical potentials and takes the form given by Eq.~(\ref{eq:newbetacondition}).

Inside a $\beta$-equilibrated NS, only the total nuclear density $\rho=\rho_n+\rho_p$ is conserved, while the neutron and proton fractions are free to adjust. This means that the individual Fermi momenta $k_{\mathrm{F}i}$ are not pinned by a fixed single-component density, but rather vary with $T$ through the modified $\beta$-equilibrium. To gain an intuitive insight into the impact of NFL behavior, we approximate the chemical potentials as $\mu_{i}=\sqrt{c^2k^{2}_{\mathrm{Fi}}+c^4 M^{\ast2}_i}\approx ck_{\mathrm{Fi}}$ at high densities and in the $T=0$ limit. Then the NFL-corrected chemical potentials reduce to
\begin{eqnarray}
&& \mu^{\mathrm{NFL}}_{i}(T)\approx\frac{h^{2}_{i}
k^{~}_{\mathrm{Fi}}}{2a_i\pi^{2}}\ln\left(\frac{T_{0}}{T}\right).
\end{eqnarray}
Substituting these equations into the new $\beta$-equilibrium relation Eq.~(\ref{eq:newbetacondition}) and then imposing the momentum conservation Eq.~(\ref{triangle}) and the electrical neutrality, we find that the nDU triggering condition is changed into
\begin{eqnarray}
\frac{k^{~}_{\mathrm{Fn}}}{k^{~}_{\mathrm{Fp}}}=\frac{h^{2}_{p}-
8\pi^{2}c \ln\big(\frac{T_{0}}{T}\big)}{\frac{5 }{2}h^{2}_{n}}< 2.
\label{newcondition}
\end{eqnarray}
The nucleon Fermi momenta now correlate with $h_{n}$, $h_{p}$, and $T$ via an isospin redistribution in the NS matter, rather than depending solely on the pressure or density. For appropriate values of these three parameters, the condition given by Eq.~(\ref{newcondition}) can always be satisfied. According to Eq.~(\ref{newcondition}), lower temperatures are more conducive to triggering DU process when the superfluid and superconducting quantum criticality are both present. Moreover, Eq.~(\ref{newcondition}) shows that stronger superconducting quantum criticality, i.e., a larger $h_p$, makes the nDU process less likely to occur, and at fixed $h_n$ a larger $h_p$ lowers the onset temperature of nDU process. Conversely, a larger $h_n$ enhances the likelihood of initiating nDU process.

\begin{figure}[htbp]
\centering
\includegraphics[width=\linewidth]{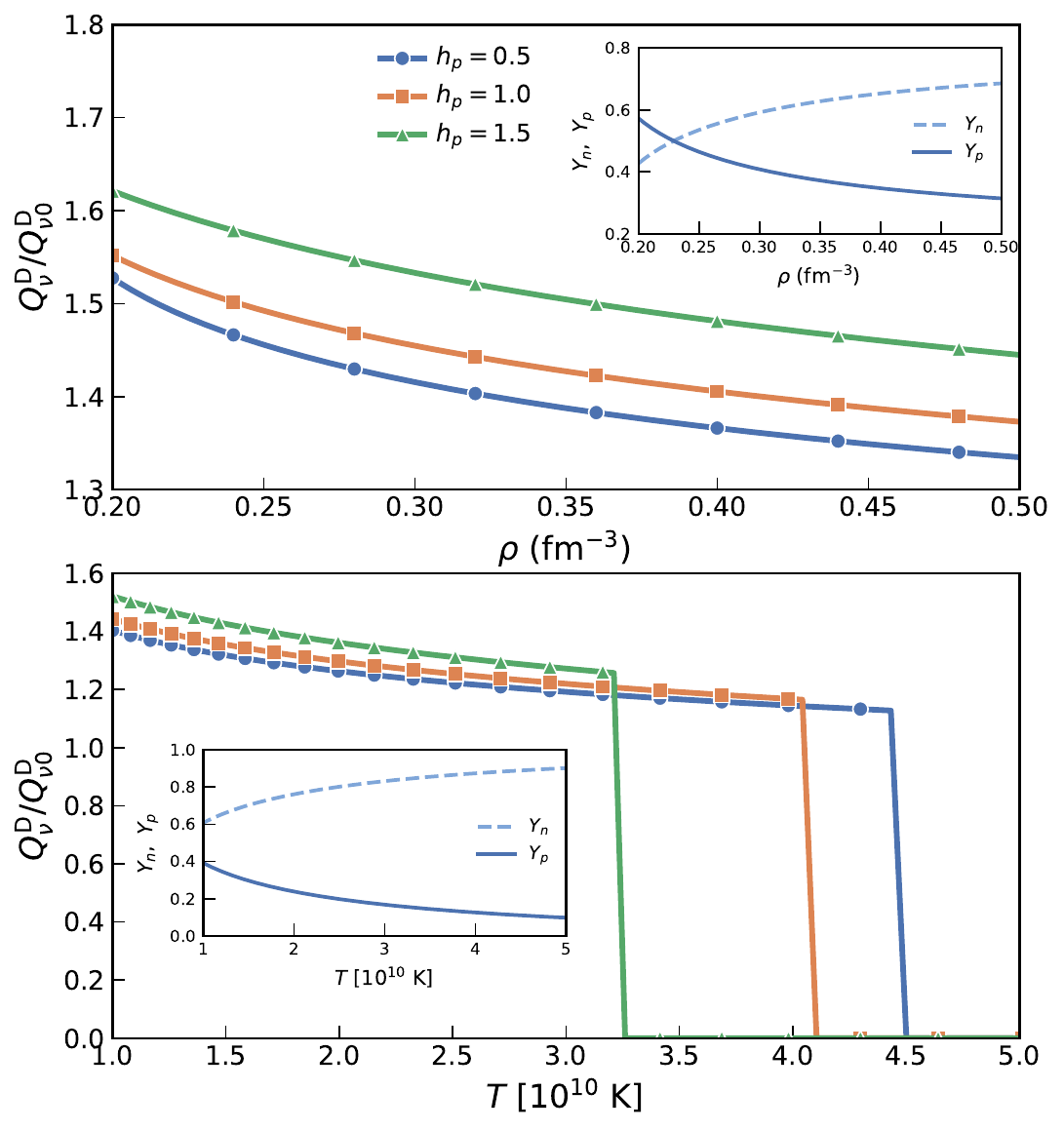}
\caption{Upper panel: $Q^{\mathrm{D}}_{\nu}/Q^{\mathrm{D}}_{\nu 0}$ versus nuclear density $\rho$ for $h_n=2.0$, $M^{\ast}_{i}=0.5M_{i}$, and $T=10^{10}~\mathrm{K}$, with $h_p=0.5, 1.0, 1.5$. The inset shows $Y_n$ (dashed) and $Y_p$ (solid) versus $\rho$ for $h_p=0.5$. Lower panel: Renormalized nDU emissivity $Q^{\mathrm{D}}_{\nu}/Q^{\mathrm{D}}_{\nu 0}$ versus temperature $T$ for $h_n=2.0$, $M^{\ast}_{i}=0.5M_{i}$, and $\rho=2\rho_0$ ($\rho_0=0.16~\mathrm{fm}^{-3}$), with $h_p=0.5, 1.0, 1.5$. The inset shows the neutron and proton fractions $Y_n$ (dashed) and $Y_p$ (solid) versus $T$ for $h_p=0.5$.} \label{fig:DUQvsT}
\end{figure}

The most important consequence arising from the modified condition of Eq.~(\ref{newcondition}) is that the nDU process occurs more readily. Under DQG/FL assumption, the proton fraction $Y_{p}$ exceeds the threshold $0.11$ only in the extremely dense core region of massive NSs. Once NFL behavior is taken into account, however, the nDU process can be triggered at much lower densities. As demonstrated in the upper panel of Fig.~\ref{fig:DUQvsT}, for a NS cooled to an appropriate internal temperature (e.g., $T = 10^{10}$~K), the proton fraction can exceed $0.11$ even at a density as low as $1.25\rho_0$, which is easily achievable in the interiors of low-mass NSs.

Besides modifying the density threshold for the nDU process, incorporating the effects of NFL behavior also substantially changes its temperature regime. To demonstrate this, we substitute Eq.~\eqref{NFLpchT} into Eq.~\eqref{eq:newbetacondition} and then perform iterative calculations at a fixed nuclear density $\rho=(k^{3}_{\mathrm{Fn}} + k^{3}_{\mathrm{Fp}})/3\pi^2$. We also calculate the renormalized nDU neutrino emissivity $Q^{\mathrm{D}}_{\nu}$ after including the NFL effects and obtain
\begin{eqnarray}
\frac{Q^{\mathrm{D}}_{\nu}}{Q^{\mathrm{D}}_{\nu0}} &=& \Big[1 +
\frac{5h^{2}_{n} M^{\ast}_n}{16\pi^2 k^{~}_{\mathrm{Fn}}}\ln
\Big(\frac{T_{0}}{T}\Big)\Big]\Big[1+\frac{h^{2}_{p} M^{\ast}_p}{8\pi^{2}k^{~}_{\mathrm{Fp}}} \ln\Big(\frac{T_{0}}{T}\Big)\Big] \nonumber \\
&& \times \Theta(k^{~}_{\mathrm{Fe}} +
k^{~}_{\mathrm{Fp}}-k^{~}_{\mathrm{Fn}}), \label{NFLDU}
\end{eqnarray}
where $\Theta(x)$ is the Heaviside step function, defined by $\Theta(x)=1$ for $x\ge0$ and $\Theta(x)=0$ otherwise, and
\begin{eqnarray}
Q^{\mathrm{D}}_{\nu0} = 4.24\times 10^{27}
\Big(\frac{k^{~}_{\mathrm{Fe}}}{1.68{\mathrm{fm}}^{-1}}\Big)
\frac{M^{\ast}_n M^{\ast}_p}{M^{2}_{n}}T^{6}_{9}
~\frac{\mathrm{erg}}{{\mathrm{cm}}^3\mathrm{s}} \nonumber
\label{FLDU}
\end{eqnarray}
is the standard emissivity of nDU process obtained under the DQG approximation \cite{Lattimer91}. As will be shown below, the logarithmic corrections in Eq.~(\ref{NFLDU}) are secondary, and the dominant NFL effect is the modified $\beta$-equilibrium condition.

The density and temperature dependence of $Q^{\mathrm{D}}_{\nu}/Q^{\mathrm{D}}_{\nu0}$ are presented in the upper and lower panels of Fig.~\ref{fig:DUQvsT}, respectively. The lower panel of Fig.~\ref{fig:DUQvsT} shows that the neutrino emissivity is enhanced with decreasing temperature. Previous DQG/FL-based studies~\cite{Yakovlev01} concluded that the nDU process is more likely to occur at higher temperatures. Our results show that including the NFL effect leads to an opposite conclusion. The reason is that the NFL effect becomes more prominent at lower temperatures, amplified by the $\ln(T_0/T)$ factor, which gradually reduces the ratio $k_{\mathrm{Fn}}/k_{\mathrm{Fp}}$ and meanwhile increases the proton fraction.

The inset of the lower panel of Fig.~\ref{fig:DUQvsT} shows that as the NS interior cools from $10^{11}$~K, the neutron fraction gradually decreases while the proton fraction increases. The critical temperature at which the proton fraction reaches its threshold $Y_p$ -- and hence the onset temperature of nDU processes -- depends on the values of $h_n$ and $h_p$. For $h_n = 2.0$ and $h_p = 0.5$, this onset occurs at $T \approx 4.5 \times 10^{10}$~K.

Conventional wisdom was that higher densities always correspond to higher proton fractions in a $\beta$-balanced NS. In the NFL regime, this trend is reversed: rising the total baryon density gradually lowers the proton fraction, which in turn progressively reduces the nDU neutrino emissivity (see the upper panel of Fig.~\ref{fig:DUQvsT}). Additionally, at a fixed density, a larger $h_p$ gives rise to a larger emissivity. These findings underscore the profound impact of NFL physics on the internal structure and composition of NSs.

\begin{figure}[t]
\centering
\includegraphics[width=1.03\linewidth]{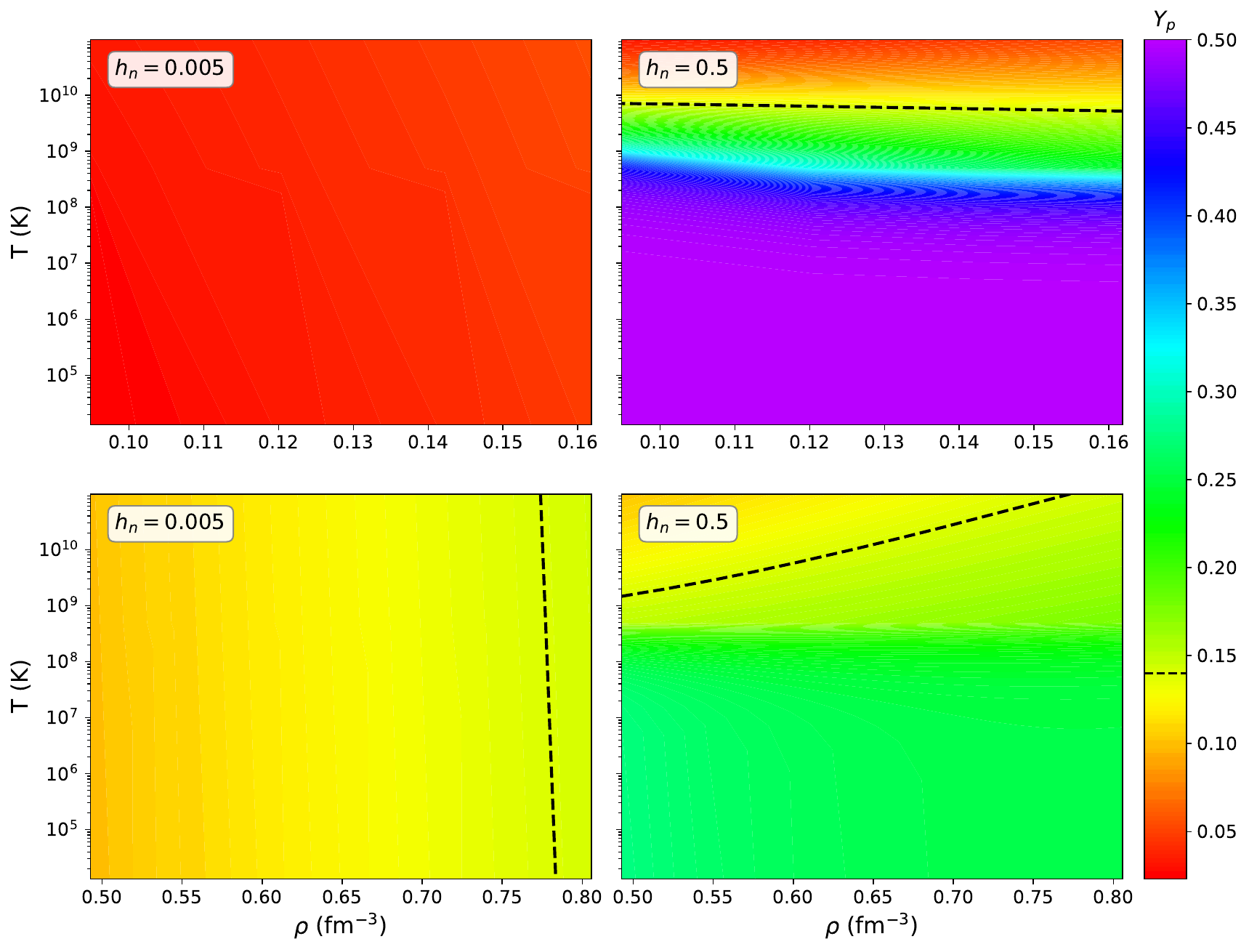}
\caption{Proton fraction $Y_p$ in the $\rho-$$T$ plane, obtained
from the NFL-corrected $\beta$-equilibrium condition
Eq.~\eqref{eq:newbetacondition} using the APR EOS with $h_p = 0.1$
fixed. Top and bottom rows show two representative density regions,
$\rho \approx (0.1-$$0.16)~\mathrm{fm}^{-3}$ and
$\rho \approx (0.5-$$0.8)~\mathrm{fm}^{-3}$, where the neutron
superfluid and proton superconducting quantum critical regions
overlap. Left and right columns correspond to $h_n = 0.005$ and
$h_n = 0.5$, respectively. The dashed black line marks the nDU
threshold $Y_p \approx 0.14$ for the APR EOS with muons included.}
\label{fig:PF2d}
\end{figure}

To relate the above general theoretical results to the realistic NS matter, we evaluate the NFL-corrected $\beta$-equilibrium condition given by Eq.~\eqref{eq:newbetacondition} using the Akmal-Pandharipande-Ravenhall (APR) EOS~\cite{Akaml98}, in which the nucleon effective masses $M_i^{\ast}(\rho)$ vary with density. The contribution of muons is included, so that for the APR EOS the nDU threshold is $Y_p \approx 0.14$ instead of the muon-free value $Y_p \approx 0.11$. Notably, the NFL effects are strongest when the superfluid and superconducting quantum critical regions overlap. In these regions, the unpaired neutrons couple to the quantum fluctuations of superfluid order parameter while the unpaired protons couple to those of superconducting order parameter. The two logarithmic corrections to $\mu_n^{\mathrm{NFL}}$ and $\mu_p^{\mathrm{NFL}}$
therefore act simultaneously, leading to most significant modification of the $\beta$-equilibrium condition. Following existing BCS calculations~\cite{Sedrakian19}, here we adopt two such overlapping regions: one at $\rho \approx (0.1-$$0.16)~\mathrm{fm}^{-3}$, where the $^{1}S_{0}$-wave superfluid and $^{1}S_{0}$-wave superconducting quantum critical regions intersect, and another at $\rho \approx (0.5-$$0.8)~\mathrm{fm}^{-3}$, where the $^{3}P_{2}$-wave superfluid quantum critical region overlaps with the $^{1}S_{0}$-wave superconducting one. Figure~\ref{fig:PF2d} shows the resulting proton fraction $Y_p$ in the $\rho$-$T$ plane for these two overlap regions, with a fixed value of $h_p = 0.1$ and two representative values of $h_n$.

When both superfluid and superconducting quantum fluctuations are weak, the system reduces to an ordinary FL. This limit corresponds to the left panels of Fig.~\ref{fig:PF2d}, where $h_n = 0.005$. The proton fraction is then basically independent of temperature, as required by the original $\beta$-equilibrium condition Eq.~\eqref{energyconservation}, and increases monotonically with increasing density. In the low-density overlap region, shown in the top-left panel, $Y_p$ remains far below the nDU threshold $Y_p \approx 0.14$. In the high-density overlap region, shown in the bottom-left panel, the threshold is crossed only at the highest densities ($\rho \geq 0.78~\mathrm{fm}^{-3}$), corresponding to $M_{\mathrm{DU}} \approx 1.97\,M_\odot$ for the APR EOS. The nDU process therefore remains confined to the dense cores of massive NSs, recovering the standard FL paradigm.

In the case of strong superfluid quantum criticality, the behavior changes qualitatively. This regime is illustrated in the right panels of Fig.~\ref{fig:PF2d}, where $h_n = 0.5$ is adopted. The proton fraction acquires a pronounced temperature dependence: as $T$ decreases, the logarithmic corrections grow and $Y_p$ is driven upward throughout the $\rho-$$T$ plane. In the high-density overlap region, shown in the bottom-right panel, the threshold $Y_p \approx 0.14$ is already exceeded well above $10^{9}~\mathrm{K}$, and $Y_p$ continues to rise as $T$ decreases further. In the low-density overlap region, shown in the top-right panel, the threshold is crossed at lower temperatures. Below $10^{8}~\mathrm{K}$, $Y_p$ approaches $0.5$ in the range $\rho \approx (0.1-$$0.16)~\mathrm{fm}^{-3}$, so the nuclear matter is nearly isospin-symmetric. In the APR EOS, this density range is readily reached inside NSs with masses of $1\,M_\odot$ or even lower. Therefore, once the enhanced NFL effects from overlapping quantum critical regions are considered, the nDU process can be activated in the outer core of low-mass NSs rather than being restricted to the dense cores of massive ones.

In summary, after including the NFL behavior arising from superfluid and superconducting quantum criticality, we demonstrate that the triggering condition for nDU process acquires qualitatively new dependence on both density and temperature.

In the density regime, the NFL behavior facilitates the occurrence of nDU process. In the conventional FL framework without considering nucleon spectral broadening~\cite{Sedrakian24, Alford24}, the nDU threshold of proton fraction is sharp, and thus the nDU process operates only in NSs with masses above a rather high threshold $M_{\mathrm{DU}}$. For example, $M_{\mathrm{DU}} \approx 1.97\,M_{\odot}$ for the APR EOS~\cite{Akaml98} and $M_{\mathrm{DU}} \approx 1.595\,M_{\odot}$ for the BSk24 EOS~\cite{Pearson18}, both of which are commonly used in thermal evolution simulations. To invoke efficient nDU emission to explain the unusually rapid cooling observed in some NSs~\cite{Marino24,Ho24,Han17,Brown18}, one needs to assume the NS mass falls within a narrow window above the threshold $M_{\mathrm{DU}}$. According to our results, incorporating the NFL effects reduces the value of $M_{\mathrm{DU}}$ and permits nDU neutrino emission to operate in NSs with much lower masses. This may broaden the stellar mass range over which nDU cooling is viable.

In the temperature regime, the logarithmic corrections intrinsic to NFL behavior progressively relax the nDU threshold at lower temperatures. This is opposite to the standard DQG/FL picture and implies that nDU emission may remain efficient deep into the late cooling stages.

Our results are semi-quantitative in several respects, and the following issues warrant further investigation. First, although the NFL effects emerge in the vicinity of QCPs, they may modify the NS EOS to some extent. A more self-consistent calculation that feeds the NFL-modified EOS back into the NS structure and cooling simulations would be desirable. Second, the present work can be straightforwardly extended to study hyperonic DU processes~\cite{Prakash92} involving nucleons (e.g., $\Lambda \to p + \ell^- + \bar{\nu}_\ell$ and $p + \ell^- \to \Lambda + \nu_\ell$). Such extensions may provide a more comprehensive understanding of NS cooling~\cite{Yakovlev01,Page06a,Potekhin15,Tsuruta23}. Moreover, it is instructive to apply the generic theoretical analysis presented in this paper to explain the unusual cooling observations of specific concrete NSs. In particular, the rapid cooling observed for some NSs within soft X-ray transients, such as SAX J1808.4$-$3658 and MXB 1659$-$29~\cite{Han17, Brown18}, which is conventionally attributed to a relatively massive star, could also occur in lower-mass accreting NSs after taking the NFL effects into account. This weakens the inference that such sources must be massive. Addressing this possibility, together with other issues listed above, will be essential for a quantitative assessment of the NFL mechanism and its observational consequences, and we leave these questions for future work.

The authors acknowledge the support by the National Natural Science Foundation of China under grants 12073026 and 12433002, the Young Scientists Fund (Category C) of the National Natural Science Foundation of China under grant 12605240, the Anhui Natural Science Foundation under grant 2208085MA11, and the China Postdoctoral Science Foundation under grant 2025M783436. The numerical calculations in this paper have been done on the supercomputing system in the Supercomputing Center of University of Science and Technology of China.

\appendix

\onecolumngrid
\section*{Supplementary materials}

\setcounter{equation}{0}
\renewcommand{\theequation}{A\arabic{equation}}

\section{Spectral function and occupation number}
We take neutrons as an example. The following calculations also apply to protons.
Near a zero-T neutron $^3P_2$ superfluid critical point,
gapless neutrons on the Fermi surface couple to the gapless quantum
fluctuations of the neutron superfluid order parameter through a
Yukawa-type interaction. Integrating out the order-parameter
fluctuations at the one-loop level yields the retarded neutron
self-energy~\cite{Zhu25}
\begin{equation}
\mathrm{Re}\,\Sigma_R(\omega)
= -C_n\,\omega\,\ln\!\frac{c_b^{2}\Lambda^{3}}{\gamma_n\,|\omega|},
\qquad
\mathrm{Im}\,\Sigma_R(\omega)
=- C_n\,\frac{\pi}{2}\,|\omega|,
\label{eq:self}
\end{equation}
with
\begin{equation}
C_n=\frac{5\,\mu_n h_n^{2}}{12\,N c_b^{2}k_{\mathrm{Fn}}c_f^{2}\pi^{2}},
\qquad
\gamma_n=\frac{h_n^{2}\mu_n^{2}\Lambda}{2k_{\mathrm{Fn}}c_f^{4}\pi^{2}}
\label{eq:coef}
\end{equation}
for the $^3P_2\,(m_J=0)$ channel. Here $\mu_n$ is the bare neutron chemical
potential, $k_{\mathrm{Fn}}$ the neutron Fermi momentum, $c_f$ and $c_b$
the neutron and order-parameter velocities, $N=2$ the neutron flavor,
$\Lambda$ the ultraviolet cutoff, and $h_n$ the effective coupling
between neutrons and the neutron $^3P_2$ superfluid order-parameter
fluctuations. The linear imaginary part implies a vanishing neutron
quasiparticle residue $Z=0$ and hence NFL behavior.

Using the linearized dispersion in Eq.~\eqref{eq:lineardisperson}, the
retarded neutron Green's function
$G_R^{-1}(\omega,\mathbf{k})=-\omega+\xi_n(\mathbf{k})+\Sigma_R(\omega)$
yields the spectral function
\begin{equation}
A(\omega,\mathbf{k})
=-\frac{1}{\pi}\frac{\mathrm{Im}\,\Sigma_R(\omega)}
{[-\omega+\xi_n(\mathbf{k})+\mathrm{Re}\,\Sigma_R
]^{2}+[\mathrm{Im}\,\Sigma_R]^{2}}.
\label{eq:spec}
\end{equation}
The neutron occupation number at zero temperature is
\begin{equation}
n(\mathbf{k})=\frac{1}{\mathcal{Z}(\mathbf{k})}
\int_{-\infty}^{0}\!d\omega\,A(\omega,\mathbf{k}),
\qquad
\mathcal{Z}(\mathbf{k})=\int_{-\infty}^{+\infty}\!d\omega\,A(\omega,\mathbf{k}),
\label{eq:occ}
\end{equation}
where the normalization $\mathcal{Z}(\mathbf{k})$ enforces the spectral
sum rule within the low-energy approximation of Eq.~\eqref{eq:self}.
In the noninteracting limit $h_n\to 0$, the self-energy vanishes and
Eqs.~\eqref{eq:spec} and~\eqref{eq:occ} reduce to
\begin{equation}
A_0(\omega,\mathbf{k})=\delta(\omega-\xi_n(\mathbf{k})),
\qquad
n_{\mathrm{F}}^{0}(\mathbf{k})=\Theta(-\xi_n(\mathbf{k})).
\label{eq:nonint}
\end{equation}
Equation~\eqref{eq:nonint} provides the noninteracting reference against which the
interaction-induced Fermi-surface smearing shown in
Fig.~\ref{fig:occupation} are contrasted.

\begin{figure}[htbp]
\centering
\includegraphics[width=0.5\linewidth]{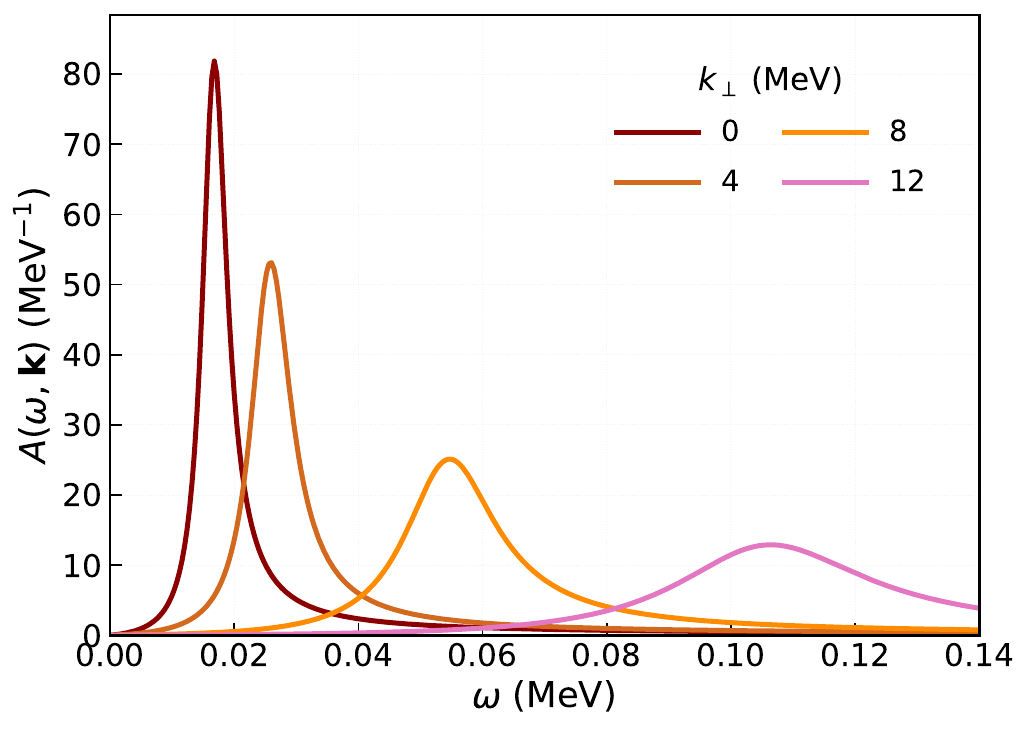}
\caption{Spectral function $A(\omega,\mathbf{k})$ in the $^3P_2$-wave neutron superfluid quantum critical region, where $k_z$ and $k_\perp=\sqrt{k_x^{2}+k_y^{2}}$ are the longitudinal and transverse momenta, respectively. The curves show $A(\omega,\mathbf{k})$ versus $\omega$ at fixed $k_z=0.05$~MeV for $k_\perp=0,4,8,12$~MeV, with $h_n=2.0$. The NFL behavior manifests itself as broadened spectral peaks.}
\label{fig:spectral}
\end{figure}

Fig.~\ref{fig:spectral} complements the occupation number
$n_{\mathrm{F}}(k_{z})$ shown in Fig.~\ref{fig:occupation} of the
main text. In the noninteracting limit (DQG), $A(\omega,\mathbf{k})$
reduces to a $\delta$ function. Once the coupling of neutrons to
superfluid quantum fluctuations is switched on, the sharp peak is
replaced by a broadened Lorentzian: as $k_\perp$ increases, the peak
shifts to higher $\omega$ and broadens progressively, with the ratio
$|\mathrm{Im}\,\Sigma_R(\omega)|/\omega$ remaining finite---in
contrast to its vanishing value in a FL. Together with the smoothed
occupation number in Fig.~\ref{fig:occupation}, this demonstrates how
the originally sharp Fermi surface with a finite $Z$ is smeared as
$Z$ approaches zero.

For Fig.~\ref{fig:spectral}, we adopt $\mu_n\approx 575$~MeV,
$k_{\mathrm{Fn}}\approx 1.68$~fm$^{-1}$ (corresponding to an
effective neutron mass $M_n^* = 0.5\,M_n$ with $M_n = 939$~MeV),
$c_f=c_b=1$, $N=2$, $\Lambda=20$~MeV, and $T=0$, treating $h_n$ as a
phenomenological coupling constant. The same parameters are used for
$n_{\mathrm{F}}(k_z)$ in Fig.~\ref{fig:occupation} of the main text.
We take the neutron $^3P_2$-wave channel as a representative case;
the same procedure applies to the other superfluid/superconducting
channels (the neutron $^1S_0$-wave and proton $^1S_0$-wave channels),
with $h_n$ replaced by the coupling for that channel
and the neutron quantities replaced by their proton counterparts.

\section{Quasiparticle residue and Kramers-Kronig relation}
The distinction between DQG, FL, and NFL behaviors is generically
captured by the quasiparticle residue $Z$~\cite{Varma,Giuliani},
\begin{equation}
Z = \lim_{\omega\to 0}
\left[1-\frac{\partial}{\partial\omega}
\mathrm{Re}\,\Sigma_R(\omega)\right]^{-1},
\label{eq:residue}
\end{equation}
where $\mathrm{Re}\,\Sigma_R(\omega)$ is inherently evaluated at zero temperature on the Fermi surface, which corresponds to $\mathbf{k}=0$ in this effective model, as the one-loop self-energy is derived from the effective Lagrangian expanded along the Fermi surface. 
Pauli's exclusion principle requires
$\mathrm{Im}\,\Sigma_R(\omega)\to 0$ as $\omega\to 0$, so that for a
self-energy of the low-energy form
\begin{equation}
\mathrm{Im}\,\Sigma_R(\omega) = -C|\omega|^{x},
\label{eq:omegax}
\end{equation}
with $C>0$ and $x>0$, the Kramers-Kronig relation~\cite{Giuliani},
\begin{equation}
\mathrm{Re}\,\Sigma_R(\omega)
=\frac{1}{\pi}\,\mathcal{P}\!\int_{-\infty}^{+\infty}
d\omega'\,\frac{\mathrm{Im}\,\Sigma_R(\omega')}{\omega'-\omega},
\label{eq:KK}
\end{equation}
gives $Z=0$ for $x\leq 1$ and $0<Z<1$ for $x>1$~\cite{Zhu25}. For the self-energy in Eq.~\eqref{eq:self}, $\mathrm{Im}\,\Sigma_R(\omega)$ is linear in $\omega$, corresponding to $x=1$.

Following standard practice to obtain a finite result, we introduce an ultraviolet cutoff $\omega_c$ and evaluate
\begin{equation}
\mathrm{Re}\,\Sigma_R(\omega)
=\frac{1}{\pi}\,\mathcal{P}\!\int_{-\omega_c}^{+\omega_c}
d\omega'\,\frac{\mathrm{Im}\,\Sigma_R(\omega')}{\omega'-\omega}.
\label{eq:KKcut}
\end{equation}
With the imaginary part from Eq.~\eqref{eq:self}, 
this integral yields
\begin{equation}
\mathrm{Re}\,\Sigma_R(\omega)
= C_n\,\mathrm{sgn}(\omega)|\omega|\,\frac{1}{2}
\left[\ln\!\left(\frac{\omega_c+|\omega|}{|\omega|}\right)
+\ln\!\left(\frac{\omega_c-|\omega|}{|\omega|}\right)\right],
\label{eq:Reint}
\end{equation}
which in the low-energy regime $|\omega|\ll\omega_c$ reduces to
\begin{equation}
\mathrm{Re}\,\Sigma_R(\omega)
\approx -C_n\,\omega\ln\!\frac{\omega_c}{|\omega|}.
\label{eq:Recut}
\end{equation}
Identifying the cutoff $\omega_c$ with the combination
$c_b^{2}\Lambda^{3}/\gamma_n$ arising from the effective model~\cite{Zhu25}, 
Eq.~\eqref{eq:Recut} reproduces the real part in Eq.~\eqref{eq:self}.
Using this real part, we obtain
\begin{equation}
\frac{\partial}{\partial\omega}\mathrm{Re}\,\Sigma_R(\omega)
=-C_n\!\left[\ln\!\frac{\omega_c}{|\omega|}-1\right],
\end{equation}
which diverges logarithmically as $\omega\to 0$. Consequently,
\begin{equation}
Z=\lim_{\omega\to 0}
\left[1-\frac{\partial}{\partial\omega}
\mathrm{Re}\,\Sigma_R(\omega)\right]^{-1}=0 .
\end{equation}
The linear imaginary part in Eq.~\eqref{eq:self}, corresponding to
$x=1$, therefore confirms $Z=0$ and the NFL nature of the nucleons
near the quantum critical point.

\section{Generalized Luttinger theorem}
In our low-energy effective model, the Fermi surface is located at $|\bf{k}|=0$, and the dispersion $\xi_i(\mathbf{k})$ vanishes there. The absolute particle number density obtained by integrating over the entire momentum space is therefore ultraviolet-divergent and not physically observable. The generalized Luttinger theorem concerns instead the change of the particle number density under interactions. At zero temperature, the change of the neutron number density is defined through the zero-frequency Green's function as~\cite{Luttinger60, Dzyaloshinskii03, Georges12}
\begin{equation}
\delta\rho_n = 2\int\!\frac{d^3\mathbf{k}}{(2\pi)^3}\,
\left[
\Theta\!\left(\mathrm{Re}\,G_R(\omega=0,\mathbf{k})\right)
-
\Theta\!~(\mathrm{Re}\,G_R^{(0)}(\omega=0,\mathbf{k}))
\right],
\label{eq:particle_number}
\end{equation}
where the factor $2$ accounts for spin and $G_R^{(0)}$ is the non-interacting retarded Green's function. From the self-energy in Eq.~\eqref{eq:self}, we have
\begin{equation}
\mathrm{Re}\,\Sigma_R(\omega\to0) = -C_n\,\omega\ln\!\frac{c_b^{2}\Lambda^{3}}{\gamma_n\,|\omega|}\bigg|_{\omega\to0} = 0.
\label{eq:self_zero}
\end{equation}
Therefore, the inverse retarded Green's function at zero frequency reduces to
\begin{equation}
G_R^{-1}(0,\mathbf{k}) = -\omega + \xi_n(\mathbf{k}) + \Sigma_R(\omega)\big|_{\omega=0} = \xi_n(\mathbf{k}).
\label{eq:GR_zero}
\end{equation}
The zero-energy singularity of $G_R$ is thus still determined by
\begin{equation}
\xi_n(\mathbf{k}) = 0,
\label{eq:singularity}
\end{equation}
which means that the Fermi surface position is not shifted by $\Sigma_R(0)$. Substituting Eq.~\eqref{eq:GR_zero} into Eq.~\eqref{eq:particle_number}, the integrand becomes
\begin{equation}
\Theta\!\left(\mathrm{Re}\,G_R(0,\mathbf{k})\right)
-
\Theta\!~(\mathrm{Re}\,G_R^{(0)}(0,\mathbf{k}))
=
\Theta\!\left(\xi_n(\mathbf{k})\right)
-
\Theta\!\left(\xi_n(\mathbf{k})\right)
=0,
\end{equation}
so the integrated particle number change vanishes:
\begin{equation}
\delta\rho_n = 0.
\end{equation}
This is the generalized Luttinger theorem in the present effective model: interactions do not change the Fermi volume, and the Fermi momentum $k_{\mathrm{F}i}$ remains fixed. The self-energy in Eq.~\eqref{eq:self} only renormalizes the quasiparticle velocity $v_{\mathrm{F}i}$; this renormalization is absorbed into the chemical potential $\mu_i(l)$ in Eq.~\eqref{NFLpchl}.

\renewcommand{\theequation}{\arabic{equation}}
\twocolumngrid
\end{document}